# RADIATION PATTERN OF PATCH ANTENNA WITH SLITS

V.Karthikeyan[1] and V.J.Vijayalakshmi[2]

[1]Department of ECE, SVS College of Engineering, Coimbatore, India
[2]Department of EEE, Sri Krishna College of Engg & Tech., Coimbatore, India

## ABSTRACT

*The Microstrip antenna has been commercially used in many applications, such as direct broadcast satellite service, mobile satellite communications, global positioning system, medical hyperthermia usage, etc. The patch antenna of the size reduction at a given operating frequency is obtained. Mobile personal communication systems and wireless computer networks are most commonly used nowadays and they are in need of antennas in different frequency bands. In regulate to without difficulty incorporate these antennas into individual systems, a micro strip scrap transmitter have been preferred and intended for a convinced divergence. There is also an analysis of radiation pattern, Gain of the antenna, Directivity of the antenna, Electric Far Field. The simulations results are obtained by using electromagnetic simulation software called feko software are presented and discussed.*

## Keywords

*Microstrip, patch, antenna, slit Directivity of the antenna, Electric Far Field*

## 1. INTRODUCTION

Microstrip patch antennas are usually used for wireless applications. A factor that influences the performance of an antenna is the structure of the patch. At present is a predictable necessitate for a packed in scrap aerial having a most favorable geometrical construction which is effortlessly to make and gives a towering aerial gain point. The corners were truncated square microstrip antenna is mainly used for single patch designs [1]. In this work, we obtained the compactness of the proposed circular polarization design because of inserting four slits of equal lengths at the corners [1]. The inserted slits reduce the size of the antenna. The methods of microwaves created them to be worn frequently in the commencement of the1990's in wireless and portable communication system. An individual be able to recognize for the reason that of their widespread tradition, micro strip scrap aerials have an assortment merits such because being slender, conformal, weightless, and small price. They carry out a few demerits together with thin bandwidth, small supremacy behavior capability, and upper hammering when arranged at superior frequencies. They are worn in the occurrence assortment of UHF (100 MHz) to 100 GHz [5]. Unlike study schemes are able to exist practical. Amongst those methods, micro strip scrap transmitters took a huge ingredient in a variety of real time uses since of their slim, conformal structures and effortlessly production effectiveness. Details of the proposed compact circular polarization design are explained. The simulated results with cadfeko software are presented and discussed [6-7].

## 2. ANTENNA DESIGN

The scrap transmitter, micro strip communication procession and earth flat surface are completed of elevated conductivity metal normally copper. The scrap is of extent *L*, breadth *W*, and located on the pinnacle of a substrate of width *h* with permittivity. The depth of the earth plane or of the micro strip is not seriously significant. Normally the elevation *h* is greatly lesser than the wavelength of procedure, although not to a great extent as lesser than 0.05 of a wavelength. A solitary nourish spherical schism procedure of the corners shortened square microstrip antenna is mainly used in single patch designs and array designs [2]. We obtained the compactness of the proposed design because of inserting the four slits of equal lengths at the corners as shown in Figure 1. These interleaved opening at the corners of the rectangle scrap consequence in dimension decrease of the thrilled fundamental-mode scrap plane.

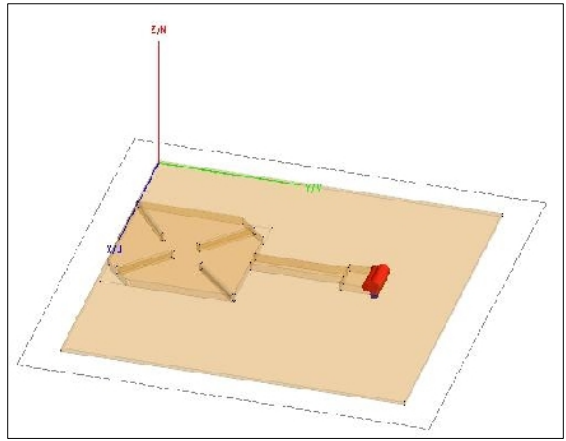

Figure 1: Microstrip patch antenna

The required specification for the circularly polarized microstrip patch antenna at 2.184 GHz is the physical requirement in order to reduce the size of the patch antennas [4]. The substrate which is duroid 5880 is chosen. The design procedure of the antennas is followed. Primarily, intend the conservative angles shortened micro strip scrap aerials at 2.184 GHz. And designed the microstrip patch antennas which have four slits. The slits are of the equal lengths. The slits are at the patch corners to achieve the size reduction [2]. The geometry of the projected micro strip aerials is shown in Figure 1, which is intended with Duroid 5880. The communication procession of 50- has a thickness 9.89 mm and extent is 10mm. The lengths of all put in slits are 18mm and breadth 1mm and with the routes of +45 point or -45 scales. The rectangle scrap has an exterior duration of 39 mm and a couple of condensed corners of magnitudes 8.5 mm × 8.5 mm [2]. The center of the agenda FEKO is based on the Method of Moments (MoM). The Method of Moments is a filled gesture resolution of Maxwell's essential equations in the incidence sphere. A benefit of the Method of Moments is with the intention of it is a foundation technique sense that merely the construction in query is damage the reputation of, not at no cost breathing space since with pasture schemes. At present is no necessitating of frontier situation to be put and reminiscence necessities balance relative to the geometry in inquiry and the necessary resolution incidence. The particular addition has been included in FEKO's Method of Moments formulation to facilitate the modeling of attractive and dielectric medium. The method of moments (Mom) or frontier constituent technique is a geometrical computational process of answering the linear fractional discrepancy equations which have been formulated as integral equations. It can be practical in a lot of areas of manufacturing and knowledge together with electromagnetic, solution technicalities, smoothness, rupture procedure and acoustics. Figure 1. The patch designed and simulated using the CADFEKO software. The radiation pattern of the Microstrip patch antenna in three dimensions is given in Figure 2. .Its E-far field polar plot is given in Figure 3. The basic patch covered now is linearly polarized since the electric field only it varies in the one direction. This polarization can be either vertical or horizontal depending on the orientation of the patch. The horizontal polarization, vertical polarization and 3D pattern requested as far field and shown in 3D view in Figure 2.

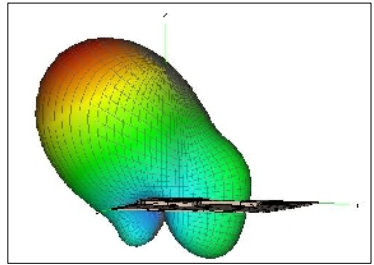

Figure 2: Radiation pattern of the Microstrip patch antenna

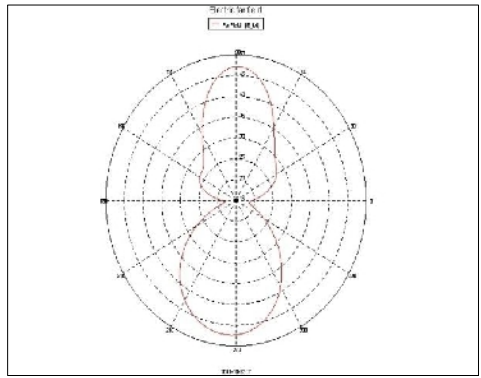

Figure 3: Electric far field for Microstrip patch antenna

## 3. RESULTS

The designed Microstrip patch antenna with four slits and a pair of truncated corners are simulated using CADFEKO for finding the electric far field radiation, gain, directivity, radiation pattern e.t.c. The electric far field of the Microstrip patch antenna is shown in figure.4.The gain of the E-far field is 75 mv. The operated frequency is 2.184 GHz.

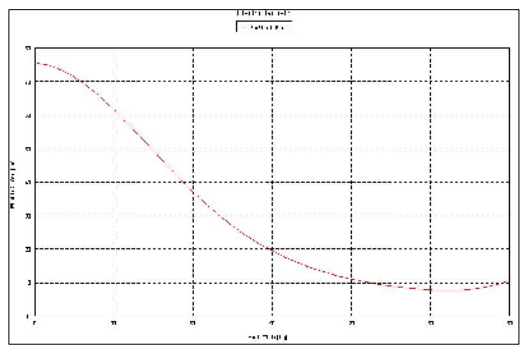

Figure 4: E-far field gain of the Microstrip patch antenna

.

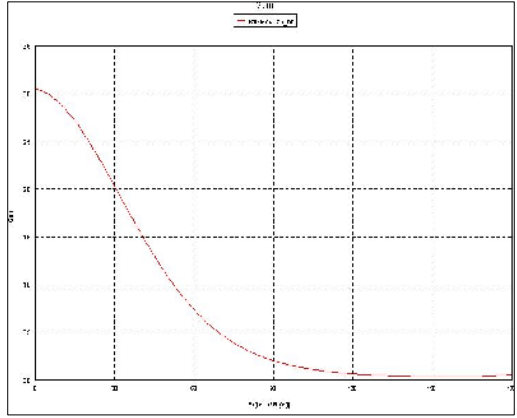

Figure 5: Directivity of the Microstrip Patch Antenna

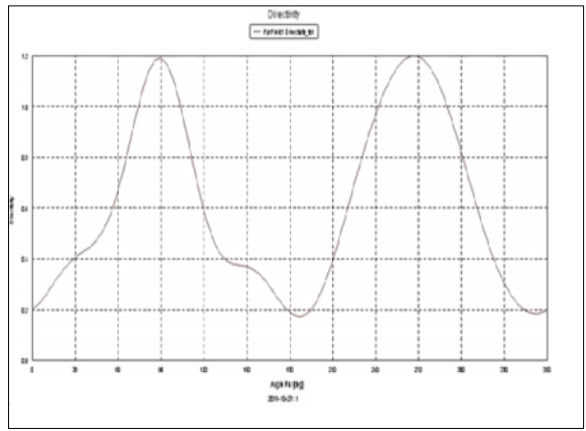

Figure 6: Gain of the Microstrip Patch Antenna

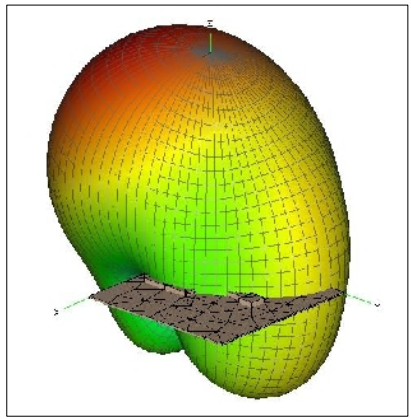

Figure 7: Radiation pattern of the Microstrip Patch Antenna

Ratio of radiated control compactness is specified by bearing to the supremacy thickness of an isotropic orientation transmitter scorching the similar entirety power. The directivity gain of the microstrip patch antenna is 1.2 dB is shown in Figure 5. The competence is distinct as the relation of the radiated power (Pr) to the contribution power (Pi). The contribution power is distorted into exuded power and exterior gesticulate power whereas a minute part is degenerated since of their performer and dielectric wounded of the resources is worn. The gain of the Micro strip scrap aerial is 32dB is shown in Figure 6. The emission prototype is a graphical representation of the comparative pasture potency transmitted as of or conventional through the transmitter. Aerial emission patterns are in use at single incidence, one divergence, and one level surface slash. The models are mostly obtainable in glacial or rectilinear appearance by means of a dB potency level. The emission mold is wide elevation.

## 4. CONCLUSION

In this work a patch antenna of microstrip which has four slits at each corner is designed and modified the antenna dimension. A patch antenna is also well-known as a rectangular micro strip antenna which is a kind of broadcasting transmitter by means of a small outline, which know how to be mounted lying on a smooth plane. It consists of a horizontal rectangular pane or piece of metal, mounted in excess of a superior pane of metal known as earth flat surface. Scrap antennas be effortless to manufacture and simply to adapt and modify. The emissions occur from discontinuities at every shortened perimeter of the microstrip communication line up. A scrap aerial is more often than not creating on a dielectric substrate. The Microstrip patch antenna with slits operated at 2.184 GHz. This frequency is used in many applications such as mobile network, wireless local area network and Bluetooth technology. The Microstrip patches antennas in terms of pattern, directivity, gain, and electric far field determined by using feko software.

**Authors**

**Prof.V.Karthikeyan** has received his Bachelor's Degree in Electronics and Communication Engineering from PGP college of Engineering and Technology in 2003, Namakkal, India, He received Masters Degree in Applied Electronics from KSR college of Technology, Erode in 2006 He is currently working as Assistant Professor in SVS College of Engineering and Technology, Coimbatore. She has about 8 years of Teaching Experience

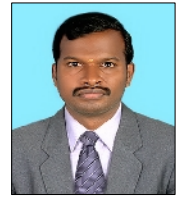

**Prof.V.J.Vijayalakshmi** has completed her Bachelor's Degree in Electrical & Electronics Engineering from Sri Ramakrishna Engineering College, Coimbatore, India. She finished her Masters Degree in Power Systems Engineering from Anna University of Technology, Coimbatore, She is currently working as Assistant Professor in Sri Krishna College of Engineering and Technology, Coimbatore She has about 5 years of teaching Experience.

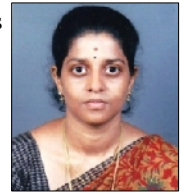